%% file: eucap_arxiv.tex
\documentclass[final, conference, a4paper]{IEEEtran}
\IEEEoverridecommandlockouts
\usepackage{xcolor}
\usepackage{amssymb}
\usepackage{graphicx}
\usepackage{amsfonts}
\usepackage{array}
\usepackage[T1]{fontenc}
\usepackage[utf8]{inputenc}
\usepackage{hyperref}
 \graphicspath{{figs/}}

\usepackage{cite}

\usepackage[cmex10]{amsmath}

\usepackage[tight,footnotesize]{subfigure}

\usepackage{url}

\usepackage{layouts}

\begin{document}
	
	\begin{titlepage}
		\begin{center}
			
			\Huge
			\textbf{A Hybrid BLE/UWB Localization Technique\\with Automatic Radio Map Creation				
			}
			
			\vspace{0.5cm}
			\LARGE
			Accepted version
			
			\vspace{1.5cm}
			
			\text{Marcin Kolakowski}
			
			\vspace{.5cm}
			\Large
			Institute of Radioelectronics and Multimedia Technology
			
			Warsaw University of Technology
			
			Warsaw, Poland,
			
			contact: marcin.kolakowski@pw.edu.pl

			\vspace{2cm}

		\end{center}
		
		\Large
		\noindent
		\textbf{Originally presented at:}
		
		\noindent
		2019 13th European Conference on Antennas and Propagation (EuCAP), Krakow, Poland, 2019
		
		\vspace{.5cm}
		\noindent
\textbf{Please cite this manuscript as:}
		
		\noindent
M. Kolakowski, "A Hybrid BLE/UWB Localization Technique with Automatic Radio Map Creation," 2019 13th European Conference on Antennas and Propagation (EuCAP), Krakow, Poland, 2019, pp. 1-4.
		
		\vspace{.5cm}
		\noindent
		\textbf{Full version available at:}
		
		\noindent
		\url{https://ieeexplore.ieee.org/document/8740154}

%
%
		
		\vfill
		
		\large
		\noindent
		© 2019 IEEE. Personal use of this material is permitted. Permission from IEEE must be obtained for all other uses, in any current or future media, including reprinting/republishing this material for advertising or promotional purposes, creating new collective works, for resale or redistribution to servers or lists, or reuse of any copyrighted component of this work in other works.
	\end{titlepage}

%
\title{A Hybrid BLE/UWB Localization Technique with Automatic Radio Map Creation}

\author{\IEEEauthorblockN{
Marcin Kolakowski, 
}                                    
\IEEEauthorblockA{
Warsaw University of Technology, Warsaw, Poland, \\
m.kolakowski@ire.pw.edu.pl}
\thanks{This work was partly supported by the National Centre for Research and Development, Poland, under Grant AAL/Call2016/3/2017 (IONIS project).}
}



\maketitle

\begin{abstract}
Localization systems intended for home use by people with mild cognitive impairment should comply with specific requirements. They should provide the users with sub-meter accuracy allowing for analyzing patient's movement trajectory and be energy effective, so the devices do not need frequent charging. Such requirements could be satisfied by employing \mbox{a hybrid} positioning system combining accurate UWB with energy efficient Bluetooth Low Energy (BLE) technology. In the paper,  such a solution is presented and experimentally verified. 
In the proposed system, user’s location is derived using BLE based fingerprinting. A radio map utilized by the algorithm is created automatically during system operation with the support of UWB subsystem. Such an approach allows the users to repeat system calibration as often as possible, which raises systems resistance to environmental changes. 
\end{abstract}


%

\vspace{7pt}
\input{tex/introduction}

\vspace{7pt}%
\input{tex/concept}
\vspace{7pt}
\input{tex/radiomap}
\vspace{7pt}
\input{tex/algorithm}
\vspace{7pt}
\input{tex/experiments}

\enlargethispage{-2.6in}%
\input{tex/conclusion}




\bibliographystyle{IEEEtran}
\bibliography{bibliography_eucap3}

\end{document}

%% file: tex/introduction.tex
\section{Introduction}
In recent years we can observe a growing market of devices and systems intended to improve elderly persons safety and quality of life. A specific example are radio localization systems intended for use by people with mild cognitive impairment in their home environment. Their main goal, aside from localizing crucial, easily lost objects (e.g. keys), is to analyze elderly persons movement and detect hazardous situations, e.g. dementia wandering \cite{linManagingEldersWandering2014}. In order for this analysis to be effective, such systems should localize the users with sub-meter accuracy. Additionally, in case of dementia sufferers, it is highly desirable that system wearables work for as long as possible without charging.

It is hard to meet those requirements using only one radio technology. Therefore, it would be reasonable to employ hybrid solution  combining accurate ultra-wideband (UWB) techniques with energy efficient Bluetooth Low Energy (BLE) technology. In the proposed system, localization would be mainly determined using a BLE subsystem, whereas UWB would be sporadically used to improve positioning accuracy.

One of the most popular localization techniques employed in Received Signal Strength (RSS) based systems is fingerprinting \cite{yiuWirelessRSSIFingerprinting2017}. It typically consists in measuring the level of signals received from system anchors and then comparing them with previously prepared radio map comprising signal levels in particular locations (signatures). 

Besides  simplicity, the main advantage of this method is that it typically does not require knowledge on  system anchors' locations and thus can be implemented with already existing infrastructure (e.g. Wi-Fi Networks at university campuses). However, fingerprinting has a significant drawback which is a big effort associated  with radio map creation. In order to provide the users with  high resolution localization, radio signatures should be collected in densely spaced points, which is time consuming. Additionally, each change in the system environment, e.g. moving furniture, disrupts propagation conditions and makes radio map outdated. Therefore, some techniques intended to make the radio map creation less time consuming have been developed.

One of the proposed solutions is crowdsourcing \cite{wangIndoorSmartphoneLocalization2016}, in which the map is created with end users participation. It is typically intended for smartphone based localization systems. User's participation can be active \cite{ledlieMoleScalableUserGenerated2011} or passive \cite{rai_zee_2012-1}. In the crowdsourcing solution, volunteer  users are provided with \mbox{a dedicated} application, which is used to collect signatures. In active crowdsourcing, it typically displays a map, on which volunteers manually mark their approximate locations. The smartphone measures levels of signals from the anchors and forms radio signatures for the marked points, which are later sent to the system server. In passive crowdsourcing, the principle is the same but users' locations are calculated automatically, for example by using smartphones' inertial sensors.

The second popular solution is to create the map in a traditional way for a lower number of points and then interpolate levels in remaining locations using propagation models. It has been shown that radio map interpolation can be successfully performed using simple exponential path loss model \cite{yiuWirelessRSSIFingerprinting2017} or Gaussian processes in more advanced solutions \cite{kumarGaussianProcessRegression2016}.

In the paper a  hybrid BLE/UWB localization technique combining both of the above approaches is presented. The radio map is created automatically during systems exploitation through concurrent BLE signal levels measurement and UWB based localization. Signatures in points, which were not visited by the system's user are interpolated using the exponential propagation model. After the complete radio map is created, the UWB subsystem is turned off and user localization is determined using BLE modules only. Radio map creation can be periodically repeated without disrupting systems usage which makes it less sensitive to environmental changes.

The rest of the paper is structured as follows. In section \ref{sec:concept} the system concept is proposed. Sections \ref{sec:radiomap} and \ref{sec:algorithm} describe the radio map creation process and localization algorithm. The results of experiments are presented in section \ref{sec:experiments}.

%% file: tex/concept.tex
\section{Localization system concept}
\label{sec:concept}
\begin{figure}[!b]
\centerline{\includegraphics[width=3 in]{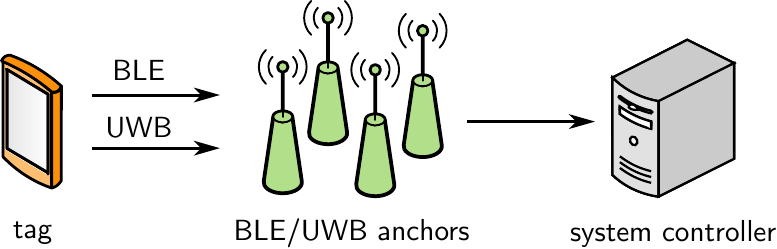}}
\caption{Proposed system functional architecture}
\label{fig:architecture}
\end{figure}

The proposed hybrid positioning system functional architecture  is presented in Fig. \ref{fig:architecture}.
The system utilizes ultra-wideband and Bluetooth Low Energy technologies. It consists of three parts:
\begin{itemize}
\item a tag worn by the user,
\item an infrastructure comprising a set of synchronized anchors including BLE and UWB modules
\item a system controller
\end{itemize}
The user is located based on  signals transmitted by the tag. The system works in two modes: calibration and normal.

In calibration mode, the radio map is  created automatically without the need for manual signatures collection. During calibration, the user wears the tag as  usual and takes a few walks,  visiting all of the rooms in his/her apartment. While the system calibration is performed the tag transmits both BLE and UWB packets concurrently. The anchors register the UWB packets time or arrival and measure the BLE signal levels and send the obtained results to the system controller. There user localizations are calculated based on UWB  time measurements and are saved with  BLE level values creating the signatures database, which will later be used to create the complete and efficient radio map of the user apartment.

In normal mode, the tag transmits  BLE packets only. The anchors measure the received signal levels and pass the results to the system controller, where the user is localized using fingerprinting and the created radio map.

%% file: tex/radiomap.tex
\section{Radio map creation}
\label{sec:radiomap}
Although typically during the system calibration several thousand signatures are registered, the created database is not fit to efficiently localize the user. First of all, the database is very large and in case of typical fingerprinting algorithms consisting in comparing measured signal levels with all of the stored signatures would be computationally expensive. Additionally, it is not guaranteed that the user, during calibration phase, visited all of the possible places in the area covered by the system. Therefore, it is necessary to process the collected signatures to create a full and computationally effective radio map.

Received signal levels in points, that were not visited by the user are interpolated using exponential path loss model which is described with the following dependence:

\begin{equation}
P_R=P_0-10\gamma \log_{10} (d/d_0) 
\label{eq:prop_mod}
\end{equation}
where:
\begin{itemize}
\item  $P_R$ - the received signal level,
\item $d$ - the distance between the anchor and the tag,
\item $P_0$ - the received signal level at the reference distance,
\item $d_0$ - the reference distance,
\item $\gamma$ - the propagation constant.
\end{itemize}

Interpolation of the radio map requires deriving $P_0$ and $\gamma$ model parameters. They are fitted for each anchor and room separately, so that wall attenuation would not introduce disturbances to the model.  Parameters are derived by minimizing the mean square error between collected signatures and values returned by the model using the Least Squares estimator employing the Levenberg-Marquardt method.

After fitting the model, in each of the rooms, a dense grid of points is created (point spacing about 10 cm). For the created mesh, the received signal levels are calculated using the fitted propagation models. The created radio map is then used to localize the user during normal system operation.

%% file: tex/algorithm.tex
\section{Localization algorithm}
\label{sec:algorithm}
User location is derived using fingerprinting.  The system employs the k-Nearest Neighbors regression algorithm \cite{dingAPWeightedMultiple2016} that works the following way. The obtained measurement results are compared with all radio map points through calculating the normalized Euclidean distance \cite{huDistanceFunctionEffect2016}:
\begin{equation}
D(x) = \sqrt{\frac{\sum_{i=1}^{m}(P_{meas}^i-P_{map}^i(x))^2}{m}}
\label{eq:euclidean}
\end{equation}
where:
\begin{itemize}
\item $x$ - the vector containing coordinates of a map point to which the measured values are compared,
\item $i$ - the  anchor number,
\item $m$ - the number of anchors,
\item  $P_{meas}^i$ - the signal level measured by anchor $i$,
\item  $P_{map}^{i }(x)$ - the signal level measured by anchor $i$ for point $x$ of the radio map.
\end{itemize}
Tag localization is computed based on k map points from which the obtained distances were the lowest. The returned localization is a weighted average of the corresponding map points coordinates:
\begin{equation}
x_L = \frac{\sum_{j=1}^{k}x_j/D_j}{\sum_{j=1}^{k}1/D_j}
\label{eq:kNN}
\end{equation}	
where:
\begin{itemize}
\item $x_j$ - a vector containing map point coordinates,
\item  $D_j$ - the normalized Euclidean distance between that map point and the measurement results.
\end{itemize}

%% file: tex/experiments.tex
\section{Experiments}
\label{sec:experiments}
The proposed technique  was tested experimentally. The main goal of the experiments was to verify whether it is possible to interpolate the radio map allowing for accurate users location determination based on a few registered calibration paths. The tests were carried out in a typical, fully furnished apartment. The system used in the study consisted of two separate UWB and BLE subsystems.

\begin{figure}[!b]
\centerline{\includegraphics[width=\linewidth]{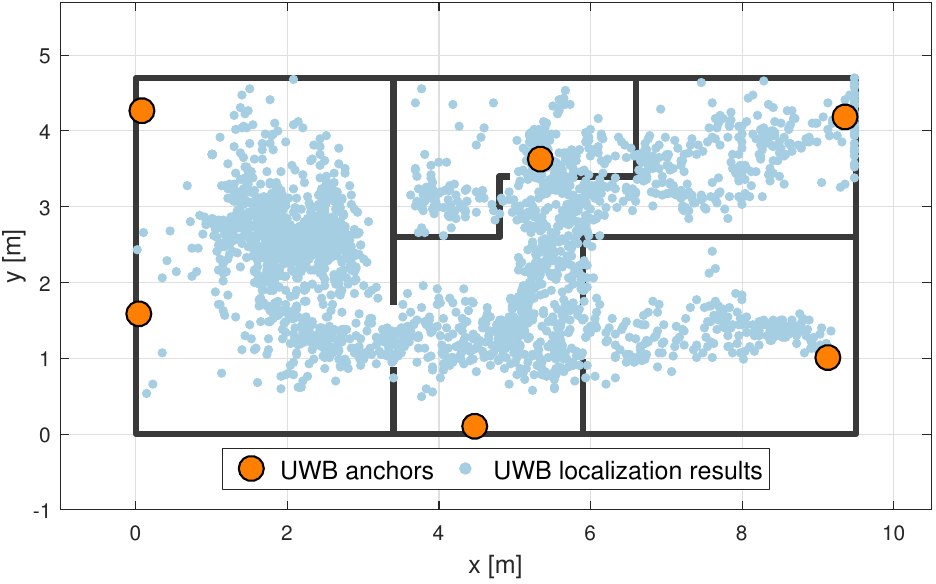}}
\caption{User's localizations obtained with UWB subsystem}
\label{fig:exp_uwb}
\end{figure}

As the UWB susbsystem, the positioning system \cite{kolakowskiUWBMonitoringSystem2017} developed within NITICS project was used. The system infrastructure consisted of six anchors which were fixed to the walls. Anchor locations and apartment layout are presented in Fig. \ref{fig:exp_uwb}. The UWB tag was fixed to the localized user's belt. It sent ultra-wideband packets with 0.16 s period. The localization was calculated using an Extended Kalman Filter based algorithm  \cite{kolakowskiUWBMonitoringSystem2017}.

The Bluetooth Low Energy subsystem was set up using Texas Instruments development kits. Four Multi-standard CC2650 LaunchPad  modules \cite{CC2650SimpleLinkMultiStandard} were used as  anchors and  CC2540 USB Evaluation Kit \cite{CC2540BluetoothLowa} was used as a tag. In the experiment, in order to make results acquisition easier, BLE transmission direction was reversed - the anchors transmitted 10 BLE packets per second and the tag acted as a receiver. The Bluetooth tag was worn at user's belt beside the UWB tag.  

The first stage of the experiment was radio map creation. It started with collecting radio signatures in the apartment. The localized user walked alongside few paths, which covered most of the apartment. The calibration process took about ten minutes, during which approximately 2500 radio signatures were collected. Since the BLE subsystem worked at a higher rate than its UWB counterpart, some of the signatures were averaged values measured for two consecutive BLE packets. User localizations derived using the UWB subsystem are presented in Fig. \ref{fig:exp_uwb}. In case of some positioning results which were located outside the apartment boundaries the results were pulled to the nearest apartment outer wall.

 \begin{figure}[!b]
\centerline{\includegraphics[width=\linewidth]{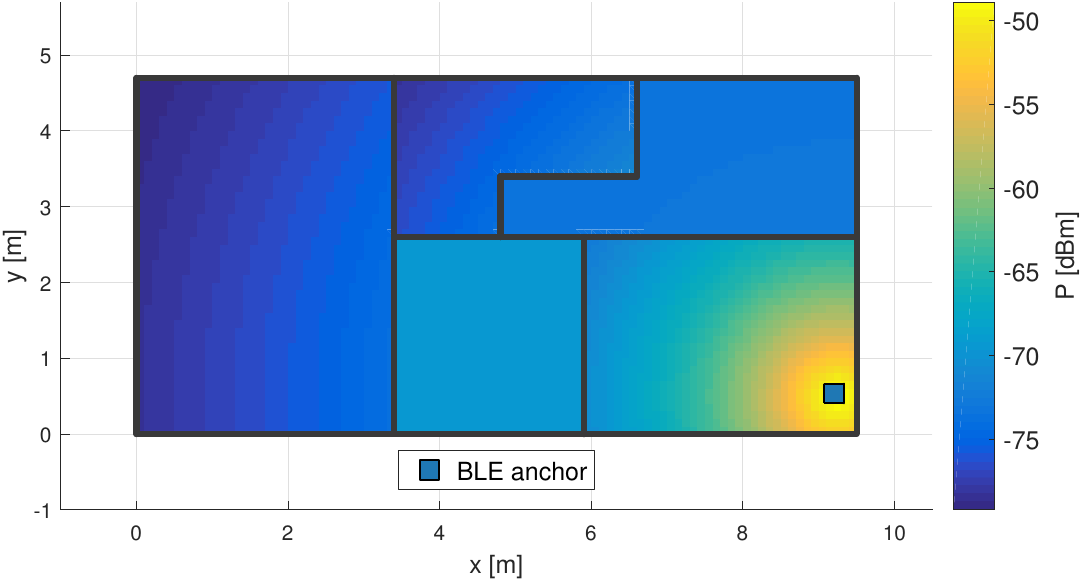}}
\caption{Estimated level of BLE signals transmitted by one of the anchors}
\label{fig:exp_model}
\end{figure}

The collected radio signatures were used to create the radio map for the whole apartment. For each of the rooms and anchors, propagation model  parameters ($P_0$ and $\gamma$ in (\ref{eq:prop_mod})) were estimated. Then the apartment plan was overlaid with a dense grid of points spaced by 10 cm. Received signal levels were interpolated in the points using fitted propagation models. Exemplary interpolated BLE signal levels for one of the anchors are presented in Fig. \ref{fig:exp_model}.

\begin{figure}[!b]
\centerline{\includegraphics[width=\linewidth]{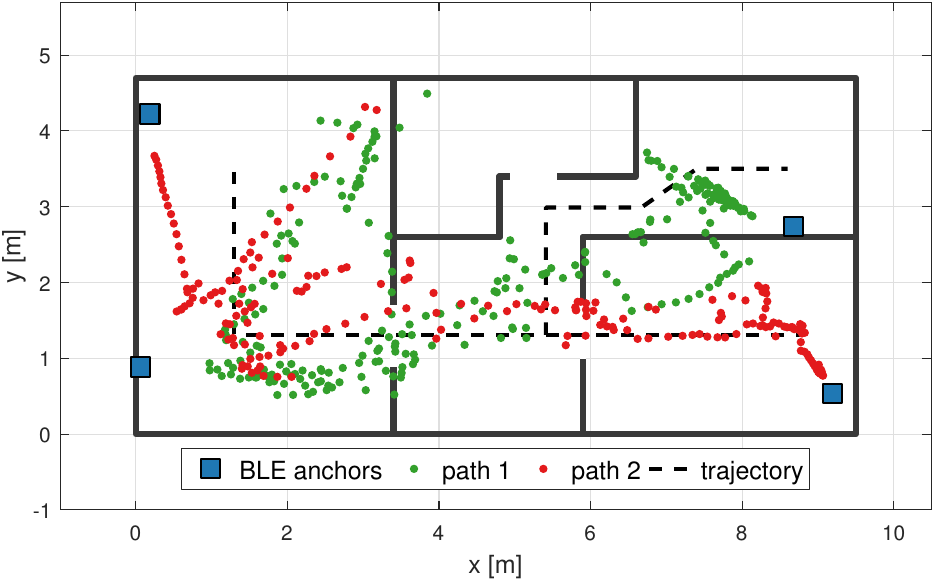}}
\caption{Localization results for two test paths}
\label{fig:exp_res}
\end{figure}

\begin{figure}[!t]
\centerline{\includegraphics[width=\linewidth]{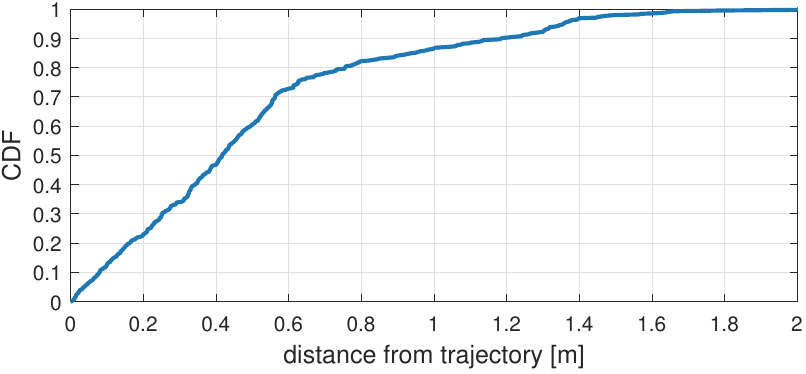}}
\caption{Estimated Cumulative Distribution Function for distance from trajectory}
\label{fig:exp_cdf}
\end{figure}

After creating the radio map, the UWB subsystem was turned off and the user was localized using BLE based fingerprinting. In order to check system's accuracy the user walked alongside few predefined paths. User's localization was derived using the k-Nearest Neighbors regressor (k=5) and the created radio map. The obtained results were additionally smoothed using moving average. As a metric of the technique performance, the trajectory error defined as a distance from reference trajectory was used. Exemplary localization results for two test paths are presented in Fig. \ref{fig:exp_res}. The Estimated Cumulative Distribution Function (CDF) curve for the trajectory error for both paths is presented in Fig. \ref{fig:exp_cdf}.

The obtained localization results are accurate enough to evaluate user's movement trajectories. The reproduced trajectories allow us to determine rooms, in which the person was located and paths covered between them. For half of the obtained results the trajectory error was lower than 65 cm. The maximum distance from the reference trajectory did not exceed 1.5 m. The accuracy of the proposed technique is at a  similar level to \cite{kumarGaussianProcessRegression2016} (average localization error  2.4 m) and better than \cite{yiuWirelessRSSIFingerprinting2017} (RMS error of few meters).

It can be noted that the above results were obtained for the BLE system consisting  of four anchors only, which is a small number compared to typical fingerprinting based systems.

%% file: tex/conclusion.tex
\section{Conclusion}
\label{sec:conclusion}
In the paper, a novel hybrid positioning technique combining ultra-wideband and Bluetooth Low Energy technologies was presented. In the proposed system, user localization is primarily derived using BLE fingerprinting, which  keeps power usage at low levels and prolong wearables use without recharging.

The radio map used in the system is created automatically by processing signatures collected during system calibration, which consists in walking along a few paths covering user's apartment. The signatures are created by connecting users location derived using the UWB subsystem with measured BLE signal levels. They  are then used to fit exponential path loss model parameters to interpolate a dense radio map covering the whole apartment. The process of radio map creation does not deprive the system of its main localization function, so it can repeated as often as necessary during normal system operation. Such a solution significantly improves system resistance to changes in the system surroundings.

The conducted experiments have shown that the created radio map allows for localization with accuracy sufficient for analyzing users movement trajectory. The presented results were achieved in a system, which infrastructure consisted of only four BLE anchors. It can be assumed that for more devices the accuracy would be higher.

The presented concept will be further developed by implementing it in a system with integrated BLE/UWB devices. Methods for interpolating radio map with other  propagation models will also be investigated.